# Frequency comb distillation for optical superchannel transmission

Chawaphon Prayoonyong, Andreas Boes, Xingyuan Xu, Mengxi Tan, Sai T. Chu, Brent E. Little, Roberto Morandotti, *Fellow OSA, Fellow, IEEE*, Arnan Mitchell, David J. Moss, *Fellow OSA, Fellow, IEEE*, and Bill Corcoran

*Abstract* — Optical frequency combs can potentially provide an efficient light source for multi-terabit-per-second optical superchannels. However, as the bandwidth of these multi-wavelength light sources is increased, it can result in low per-line power. Optical amplifiers can be used to overcome power limitations, but the accompanying spontaneous optical noise can degrade performance in optical systems. To overcome this, we demonstrate wideband noise reduction for comb lines using a high-Q microring resonator whose resonances align with the comb lines, providing tight optical filtering of multiple combs lines at the same time. By distilling an optical frequency comb in this way, we are able to reduce the required comb line OSNR when these lines are used in a coherent optical communications system. Through performance tests on a 19.45-GHz-spaced comb generating 71 lines, using 18 Gbaud, 64-QAM sub-channels at a spectral efficiency of 10.6 b/s/Hz, we find that noise-corrupted comb lines can reduce the optical signal-to-noise ratio required for the comb by ~ 9 dB when used as optical carriers at the transmitter side, and by ~ 12 dB when used as a local oscillator at the receiver side. This demonstration provides a method to enable low power optical frequency combs to be able to support high bandwidth and high-capacity communications.

## I. Introduction

OPTICAL frequency combs are unique light sources consisting of precisely equidistant frequency lines, that have enabled many applications ranging from spectroscopy [1]–[3] to neural networks [4], [5], and coherent optical communications systems [6]. Although frequency comb sources are commercially available, these benchtop solutions may not be suitable for many practical applications because of their size, weight, power requirements and cost. These limitations can potentially be overcome by the use of miniaturized 'microcombs' [7]–[9] that have proven to be attractive solutions for a wide range of applications, including microwave photonics [10], [11], optical communications [12]–[14] and others.

In optical communications systems, the individual optical frequency comb lines can be deployed either as carriers at the transmitter, or as a local oscillator (LO) at the receiver, to support superchannel transmission. High capacity optical superchannels have the potential to serve future data capacity requirements in existing fiber links, where a high spectral efficiency is critical to increase the capacity and optical bandwidth of systems over standard single mode fiber [14]–[16]. In these superchannel systems, optical frequency combs offer an attractive alternative to using a large number of conventional external cavity lasers (ECLs) to support many individual sub-bands [12]–[14]. Furthermore, optical frequency combs also offer advantages in terms of simplifying any signal processing, while at the same time reducing the system power consumption and cost [17]–[19]. These benefits scale with the number of comb lines generated, to cover a wider bandwidth for communications.

Wide bandwidth combs can be generated via various mechanisms, such as nonlinear frequency conversion ([10], [12], [20]–[23]), the electro-optic effect ([24]–[26]) of high speed modulators, and mode locked lasers ([27], [28]). Since majority of combs convert the power of the initial laser line to the comb lines, a larger number of generated comb lines results in a lower power per line [17]. In addition, micro-combs generally tend to have a nonuniform spectral power distribution which can be undesirable for some applications. Hence, if the comb spectrum is flattened, by e.g. filtering with a gain-flattening filter or a waveshaper [14], the average power per line [6], [12]–[14] will be further reduced.

This work was supported in part by the Australian Research Council (ARC) under Grant DP190102773.

C. Prayoonyong and B. Corcoran are with Photonic Communications Laboratory, Dept. Electrical and Computer Systems Engineering, Monash University, Clayton, VIC 3800, Australia (e-mail: chawaphon.prayoonyong@monash.edu; bill.corcoran@monash.edu).

A. Boes and A. Mitchell are with School of Engineering, RMIT University, Melbourne, VIC 3001, Australia (e-mail: andreas.boes@rmit.edu.au; arnan.mitchell@rmit.edu.au).

M. Tan and D.J. Moss are with the Optical Sciences Centre, Swinburne University, Hawthorne, VIC 3122, Australia (e-mail: mengxitan@swin.edu.au; dmoss@swin.edu.au).

X. Xu was with Centre for Micro-Photonics, Swinburne University, Hawthorne, VIC 3122, Australia. He is now with the Photonic Communications Laboratory, Dept. Electrical and Computer Systems Engineering, Monash University, Clayton, VIC 3800, Australia (e-mail: mike.xu@monash.edu)

S. T. Chu is with Department of Physics, City University of Hong Kong, Tat Chee Ave, Hong Kong, China (e-mail: saitchu@cityu.edu.hk)

B. E. Little is with Xi'an Institute of Optics and Precision Mechanics of the Chinese Academy of Sciences, Xi'an, China (email: b.e.little@ciompxian.edu.cn)

R. Morandotti is with INRS – EMT, Varennes, Quebec J3X 1S2, Canada & Adjunct with the Institute of Fundamental and Frontier Sciences, University of Electronic Science and Technology of China, Chengdu 610054, China (e-mail : morandotti@emt.inrs.ca)

To compensate for this, one can amplify the comb, typically with Erbium-doped fiber amplifiers (EDFAs). In this case, however, optical noise stemming from amplified spontaneous emission (ASE) can contaminate the amplified comb lines. In optical communications systems, this broadband noise could degrade the performance of comb lines when they are used as either carriers or LOs, particularly for high modulation formats such as 64QAM where high signal-to-noise ratios are required.

There are a variety of approaches used to purify the amplified comb lines by reducing the wideband optical noise surrounding the narrowband comb lines, including techniques such as optical injection locking [29]–[31] and stimulated Brillouin scattering (SBS) [32], [33]. However, these approaches involve limitations due to size, power consumption, and/or practical limits to their optical frequency range.

Here, we extended of our work in [34] where we adapted an approach employed in spectroscopy and microwave photonics [35], to simultaneously filter a large number of amplified EO generated frequency comb lines with a high-quality (Q) factor MRR (Q ~ 5 x$10^5$), for applications to optical communications systems. In this work, we assess benefits of at both transmitter and receiver sides in terms of signal quality ($Q^2$), bit error ratio (BER) and generalized mutual information (GMI). Our results show that the performance of the system based on distilled comb lines is significantly better than that based on noisy (amplified, unfiltered) lines. The MRR filter reduces the system noise, resulting in a reduction in the required optical signal-to-noise ratio (OSNR) for comb lines by ~ 9 dB and ~ 12 dB when employed as carriers or local oscillators, respectively. This work demonstrates that passive filtering with a high Q MRR can significantly improve the performance of systems based on low-power optical frequency combs, thus providing an attractive path towards improving the performance of power constrained combs, as is typically the case with micro-combs.

## II. Proof of concept for comb distillation

Amplification of low power frequency combs by an EDFA can contaminate the comb lines with optical noise, degrading the performance of optical communications systems. This noise can have a particularly degrading effect on the overall communication system performance where advanced modulation formats (e.g. higher order QAM) are used. It has been shown that MRRs with moderate Q factors can provide an attractive and compact approach to filter optical frequency combs, with the drop port of a double-bus ring providing a periodic inverse-notch profile [36]. Here, we propose and demonstrate the use of a high Q factor (Q ~ 5 x $10^5$) MRR [37], [38] with a bandwidth of < 500 MHz to act as a narrowband filter for each comb line. In order to effectively distill comb lines simultaneously, it is crucial to accurately match the line spacing of the microcomb to the FSR of the MRR, so that the comb lines are allowed to pass, while broadband noise is rejected. Fig. 1 (a)-(b) illustrates the filtering achieved by the MRR, where comb lines are shown in black with a high noise floor (red), along with the resonator response depicted by the blue curve. Also shown is the output at the drop port of the MRR with a small portion of noise inside the resonance bandwidth around the comb lines remaining after distillation.

To investigate the response of the MRR, we loaded optical noise to the input port and measured the output signal at the drop port of the device. We use a 150 MHz-resolution optical spectrum analyzer (OSA, *Finisar* Waveanalyzer 1500S), and the loaded noise and output profiles are shown in Fig. 2 (a). From the measurement of through and drop port losses, the ~ 12.5 passband drop-port loss and ~ 11.5 dB off-resonance through-port loss can be mainly attributed to coupling losses between the pigtailed input fibers and the on-chip waveguides. The platform we used has the propagation loss on the order of 0.05-0.07 dB/cm [37], we therefore expect << 0.1 dB total propagation loss, causing us to infer an average per facet fibre-chip coupling loss of about 5.5-6 dB. The difference between drop-port on-resonance insertion loss and through port off-resonance loss is ~ 1 dB, which could represent the loss over the MRR. Fiber-chip coupling losses on the platform we used have been demonstrated to be as low as 3 dB [38], and so we believe that the overall insertion losses of future devices could easily be reduced to < 4dB, and maybe lower with further development. The ring we used was not designed for the comb filtering operation we used it for, and further investigations of optical ring design for this application would be required to understand the trade-offs between MRR response design and distillation performance. This highlights the scope of this technique to be further improved, to provide even greater gains in effective OSNR than we demonstrate here.

To gain more insight how the MRR filters comb lines, we measured the passband loss at resonance peaks and stopband loss as shown in Fig. 2 (b). This also confirms the flatness of passband loss (~ 12.5 ± 1 dB) and stopband loss (~ 40 ± 0.4 dB), which gives rise to extinction ratio of ~ 27.5 ± 1 dB. Therefore, it is important to match the central wavelength and line spacing of the comb with a resonance peak and FSR of the MRR, respectively, in order to minimize loss for the comb lines while attenuating the noise. While this is readily accomplished with EO combs, for micro-combs the spacing of the MRRs used for the comb and the passive filtering would need to be closely

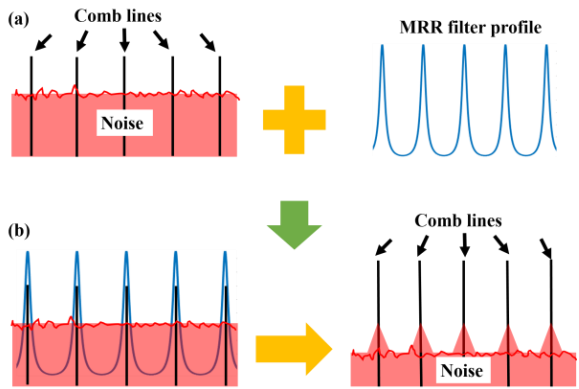

Fig. 1 (a) Comb lines with high noise floor (black and red, left) and the drop port filter profile (blue, right), (b) Filtering by microring resonator. The filter profile superimposed on the noisy comb (left) produces a noise reduced "distilled" output (black and red, right).

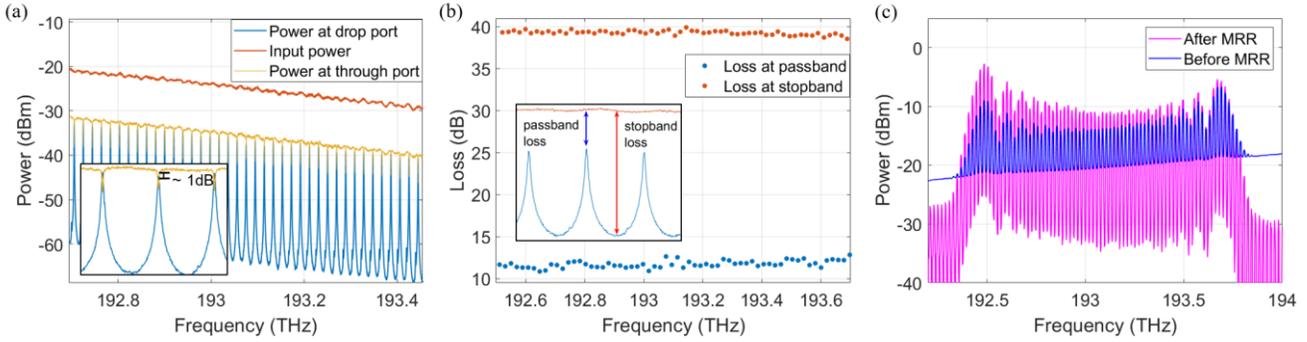

Fig. 2 (a) Noise profile at the input port of MRR (red) and corresponding output at the drop port (blue) and through port (yellow), (b) Passband and stopband loss with an inset showing the loss in the drop port output profile, (c) Spectra for amplified (blue) comb before being distillation and distilled (magenta) EO comb.

matched during fabrication, with further fine tuning being performed possibly with thermal control. Fabrication of the two matched MRRs may be simplified if they were on the same chip.

To demonstrate distillation, we generated an EO frequency comb using a set of phase modulators cascaded with an intensity modulator [25], before transmitting it through the MRR. Initially, we swept the frequency of a CW optical signal launched into the MRR in order to find a resonance peak, located at 193.089 THz. We then set the CW laser to this frequency to form the central frequency of the EO comb. Next, to align the EO comb lines to the MRR resonances, we searched for the FSR of the MRR by launching optical noise into the MRR before measuring the response with the 150 MHz-resolution OSA. We measured the FSR to be ~19.5 GHz and, by fine tuning the EO comb spacing, further resolved this to an FSR of ~19.45 GHz. Note that the MRR is polarization and temperature dependent, and so a polarization controller and thermistor with a thermo-electric cooler were used to align the EO comb polarization with the MRR, and stabilize the MRR frequency response, respectively.

Once the EO comb was adjusted to match both the resonance peaks and FSR of the MRR, we then investigated the effect of comb distillation by measuring the spectrum of the comb before and after distillation. Noise loading was achieved by attenuating the EO comb to low power followed by amplification with high gain to contaminate with the optical noise before transmitting to the MRR. We note that this approach could emulate the situations where low power or wideband combs, e.g. [14], [39], [40], are amplified to bring their line power to meet power requirements of the communication systems. Fig. 2 (c) shows two optical spectra measured with 12.5 GHz (0.1 nm) resolution, with an amplified (or noisy) comb before distillation, depicted in blue and a distilled comb in magenta, for an overall comb bandwidth of ~ 1.3 THz. It can be seen that the noise floor for the amplified comb is at ~ -20 dBm whereas its power per-line is ~ -10 dBm on average, resulting in a per-line OSNR of ~ 10 dB. The slight tilt in the noise floor of noisy

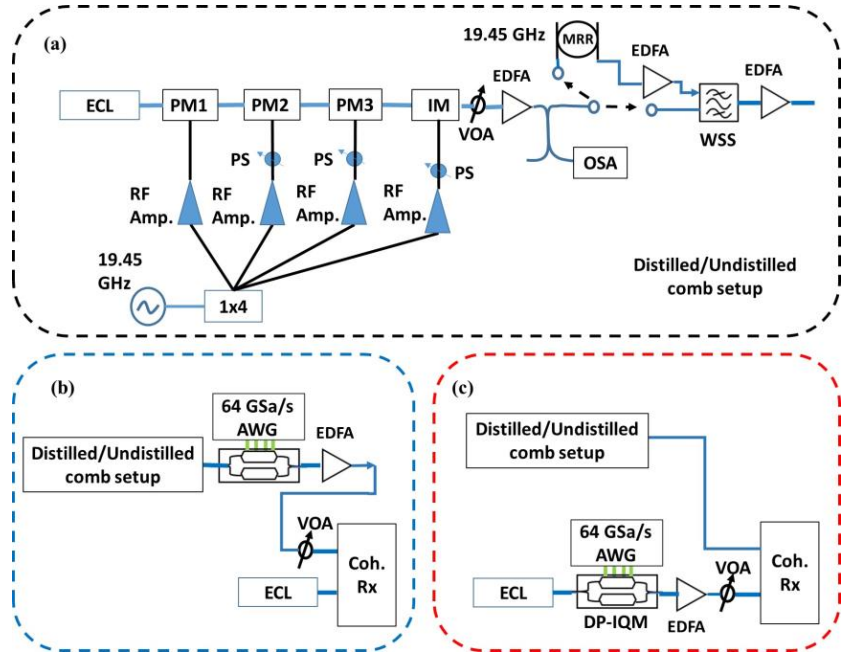

Fig. 3 (a) Distilled/undistilled comb source, and demonstration when using the comb source as (b) Carriers, (c) LOs; here ECL: External cavity laser, PM: Phase modulator, IM: Intensity modulator, PS: Phase shifter, OSA: optical spectrum analyzer, WSS: Wavelength selective switch, VOA: Variable optical attenuator, MRR: Microring resonator, AWG: Arbitrary waveform generator, DP-IQM: dual-polarization I/Q modulator, Coh. Rx.: Coherent receiver comb (blue) is caused by gain tilt of the EDFAs [41]. We think that gain tilt with opposite direction may contribute to the equalized distilled comb (magenta).

After distillation via the MRR, the out-of-band noise, as measured with a 12.5 GHz resolution, was significantly reduced compared to the comb lines, highlighting the reduction in passed noise bandwidth by the MRRs. Comparing the power of the comb lines to the noise amplitude (i.e., the OSNR) at either end of the spectrum, we observed that this was improved by approximately 20 dB, which is expected based on the difference between the ring resonance linewidth of ~ 150 MHz and the optical spectrum measurement resolution of 12.5 GHz. This indicates that broadband noise can be significantly reduced by MRR filtering. In the context of an optical communications system, this can be used to improve the quality of optical carriers amplified from low power comb sources at the transmitter and receiver.

### III. System Configurations

To investigate the benefits of comb distillation in both the transmit and receive configurations, and experimental setup is used as shown in Fig. 3 (a)-(c). An EO frequency comb was generated by launching an ECL laser with the power of 13 dBm (Keysight N7714A with laser linewidth < 100 kHz) to three cascaded phase modulators ($V\pi$ ~ 4 V at 20 GHz) together with one intensity modulator ($V\pi$ ~ 6 V at 20 GHz) as shown in Fig. 3 (a). This resulted in EO comb possessing the bandwidth identical to those in Fig. 2 with the power of -7 dBm (comb OSNR > 45 dB). Here, the three phase modulators were used to broaden the optical spectrum while an intensity modulator was used to flatten the spectrum [25]. The modulators were driven with amplified RF signals generated by a signal generator with the power of 6 dBm before being split to RF amplifiers (CTT Inc., APW/265-3335 amplifiers (30 dB gain) for the phase modulators).

To emulate amplification of low power combs, we attenuated the EO comb by varying its total power from -10 to -30 dBm. Then, we amplified this with an EDFA with the nominal output power set to 21 dBm. Although the actual power after this stage could not reach the nominal figure, noise added from this practice was adequate to degrade per-line OSNR. To measure the OSNR after this stage, 50% of the comb power was coupled to an OSA (Ando AQ6317) with a 0.1 nm resolution. From now, we will use $OSNR_{line}$ to represent the per-line OSNR measured prior to distillation or bypass paths. The remaining signal was then transmitted through the MRR via a pigtail fiber, with the drop port connected to a second amplifier, followed by a wavelength selective switch (WSS, *Finisar* Waveshaper 4000S) set with an optical bandwidth of 10 GHz, so that the comb line of interest could be used by the optical communication system. The system was configured to also allow the noise-loaded EO comb to bypass the MRR to enable a performance comparison. The comb lines were then either deployed as carrier lines at the transmitter with the ECL laser used as LO lines or vice versa (Fig 3 (b) and (c)). At this stage, the EDFA after the WSS was exploited to stabilize line power delivered to the communication systems. The carrier power before the IQ modulator was set to 11 dBm and increased to 14 dBm when used as a LO for coherent detection.

For the optical communications system, a 23-GHz bandwidth dual-polarization (DP) IQ modulator (Sumitomo 100G) was driven with a 64-GSa/s, 25-GHz bandwidth arbitrary waveform generator (AWG, Keysight M9505A) with a 2.5% root-raised cosine shaped, 64-QAM signal at 18 Gigabaud (GBd) to emulate superchannel transmission with this 19.45 GHz spaced EO comb. The optical signal was amplified before mixing with the corresponding LO line at the 25-GHz bandwidth coherent receiver (*Finisar* CPRV1220A). We note that the received data band OSNR ($OSNR_{data}$) is > 37 dB when modulated with the clean ECL line, hence optical filtering is not necessary to remove noise from the receiver front-end amplifier. In addition, we also employed the variable optical attenuator (VOA) in front of the receiver to adjust the signal power to occupy full ADC range of the oscilloscope. After coherent mixing, the signal was then sampled by an 80 GSa/s oscilloscope having a 33 GHz bandwidth (Keysight, DSOX95004Q). Next, the digitized signal was processed with a set of DSP algorithms. The DSP chain starts from IQ imbalance with Gram-Schmidt process and frequency offset compensation (peak search algorithm) [42] before resampling to 2 Sa/symbol to emulate the conventional anti-aliasing processes [43]. The signal was then matched with the RRC filter (the same roll-off factor as the transmitter side) to maximize the signal of interest [42] before being frame-synchronized using a short BPSK training overhead. Subsequently, we equalized the signal with the two-step 41-tap equalizer (Least mean square (LMS) algorithm for pre-convergence of following multi-modulus algorithm [44]). Finally, training-based phase compensation was applied to the equalized signal [45]. Following these DSP steps, the signal quality ($Q^2$) which is related to error vector magnitude (EVM) as $Q^2 = 1/EVM^2$ (i.e. equivalent to SNR in the Gaussian approximation [46]), and bit error ratio (BER) were calculated. In addition, generalized mutual information (GMI) was also calculated with the *calcGMI.m* script [47] to convey information rates.

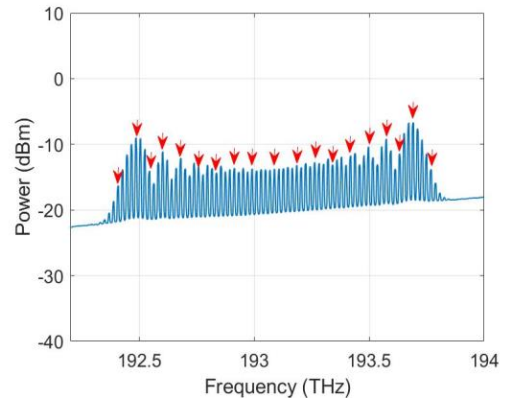

Fig.4 Electro-optic frequency comb after attenuation and amplification with the red arrows representing the lines used in the experiment.

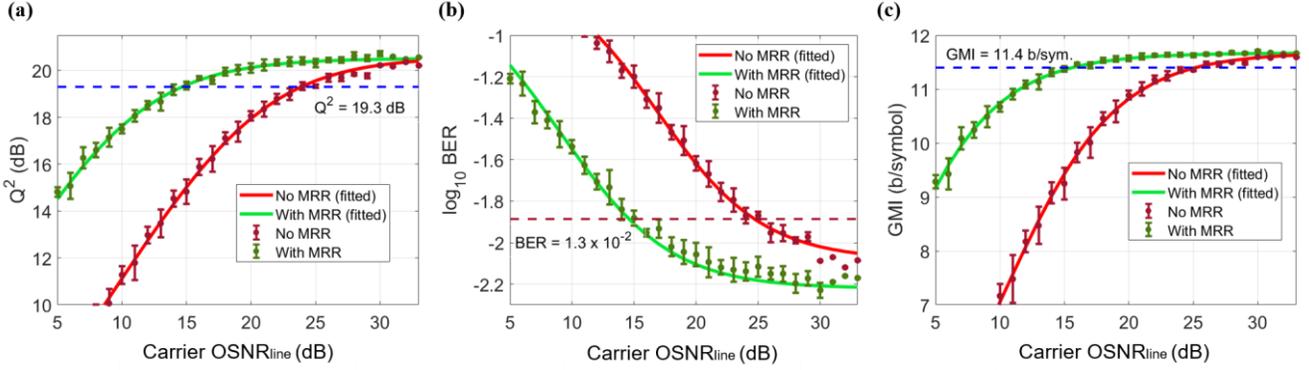

Fig. 5 Performance when noisy and distilled comb lines were employed as carriers in terms of (a) Signal quality ($Q^2$), (b) BER and (c) GMI with dashed lines indicating limits of the 5% reduction of information rates in terms of the particular metrics. Here, the red and green marks indicate undistilled and distilled performance, respectively. The points and error bars show the average and standard deviation of different tested comb lines for specific carrier $OSNR_{line}$ values. The red and green lines are guides for the eye for the undistilled and distilled lines, respectively.

## IV. DEMONSTRATION OF COMB DISTILLATION FOR CARRIERS

### A. Performance in terms of $Q^2$, BER, and GMI

In this section, we investigate and compare performance of the EO comb lines when used as carriers for superchannel transmission, both before and after distillation with the MRR. Here, the comb lines tested in the experiment are marked by the red arrows as shown in Fig. 4

Fig. 5 shows the $Q^2$, BER and GMI plotted against comb line OSNR when the comb lines were used as optical carriers for modulation. Since all tested comb lines possessed similar results, scatter points and error bars were deployed to present overall performance with deviations. Here, we also introduce the fitted curves to convey general trends. The performance is shown in red dots and fitted curves for the comb without distillation, and in green dots and fitted curves for the comb filtered with MRR.

As seen in Fig. 5 (a), the quality factor, $Q^2$, of the noisy comb (red) increases linearly before being effected by transceiver noise limits that limit performance to around $Q^2 \sim 20$ dB at an $OSNR_{line}$ of > 30 dB, suggesting that the $OSNR_{line}$ above 25 dB is sufficient for the optical carriers in our system to support penalty-free operation. This is in contrast to $Q^2$ of the filtered comb (green) where the metric reaches the transceiver noise limits ($Q^2 \sim 20$ dB) when an $OSNR_{line}$ of > 15 dB, reducing the required $OSNR_{line}$ for attaining the limits by $\sim 10$ dB compared with the noisy comb. Moreover, it is seen that filtering by MRR can provide significant improvement in $Q^2$ up to 7 dB at low $OSNR_{line}$, i.e. where the optical noise is dominant. However, once $OSNR_{line}$ increases, the improvement is less noticeable since the optical noise is comparable to the transceiver noise, which provides an upper limit to performance in the system. To benchmark the performance of the EO comb lines, we measured the required $OSNR_{line}$ at $Q^2 = 19.3$ dB, corresponding to 0.6 b/symbol or a 5% reduction in GMI for dual polarization 64QAM. At this indicative limit of $Q^2 = 19.3$ dB as illustrated with a blue dashed line, we find that the required $OSNR_{line}$ is in a range of 22-25 dB and 13-15 dB for noisy and distilled comb lines, respectively. This agrees with the 10 dB $OSNR_{line}$ reduction mentioned previously.

When investigating the effect of distillation on BER (Fig. 5 (b)), we see that the BER mirrored the trends in $Q^2$ - dropping and hitting an error floor at a BER below $10^{-2}$ when the $OSNR_{line}$ was > 30 dB and > 20 dB for un-distilled and distilled combs, respectively. The threshold of $Q^2 = 19.3$ dB in the $Q^2$ plot translates into a BER of $1.3 \times 10^{-2}$, where we see that the required $OSNR_{line}$ was similar to that shown in the $Q^2$ plot for both scenarios, corresponding to $\sim 10$ dB $OSNR_{line}$ penalty reduction at the threshold. This indicates that the noise could

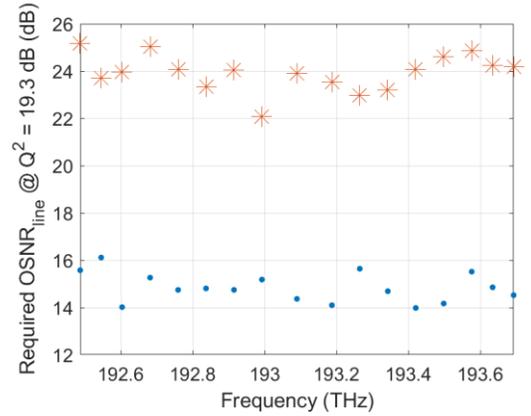

Fig. 6. Required $OSNR_{line}$ at $Q^2 = 19.3$ dB, for both bypassed and distilled carriers against frequency of comb lines.

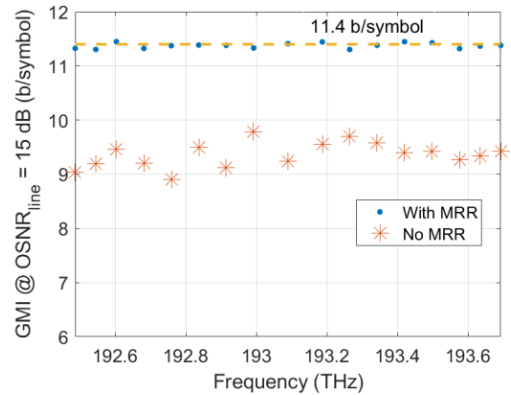

Fig. 7 GMI plot for distilled and bypassed carrier lines at $OSNR_{line}$ of 15 dB with the limit of GMI = 11.4 b/symbol as indicated with the dashed line against frequency of comb lines

still be considered a Gaussian distributed noise field contribution after distillation.

For the GMI plot in Fig. 5 (c), the GMI from both scenarios exhibit similar trends to $Q^2$, levelling off at ~11.6 b/symbol. Similar to improvement in $Q^2$, the gain in information rates is noticeable (up to 3 b/symbol) when distillation was applied to the lines at low $OSNR_{line}$, but negligible at high $OSNR_{line}$ due to the transceiver noise. With the reference for $Q^2$ and BER as given in Fig. 5 (a), (b), this relates to the GMI limit of 11.4 b/symbol regarded as a 5% reduction of the ideal achievable information rates of 64QAM at 12 b/symbol. Again, the $OSNR_{line}$ in the two cases that is required to achieve this limit also coincides with that needed for the $Q^2$ and BER.

From the results presented in Fig. 5, we notice that if the $OSNR_{line}$ of the tested lines is above 25 dB, the performance can exceed our benchmark of a 5% reduction in GMI. Therefore, we could regard $OSNR_{line}$ = 25 dB as the required carrier $OSNR_{line}$ for our EO comb to achieve the benchmark without narrowband filtering for lines, which can be used to inform us of power requirements for optical frequency combs when used as carriers in optical communication systems.

### B. Performance at the indicative threshold of each comb line

To gain further insight into the performance of the system in terms of $OSNR_{line}$, for $Q^2$ = 19.3 dB, we interpolated the $Q^2$ values from each comb line with and without distillation and solved the fitting equations for the required $OSNR_{line}$. Fig. 6 shows that required $OSNR_{line}$ is generally flat, with small fluctuations, for all frequencies, which implies frequency independence of MRR filtering on the amplified EO comb. The average required $OSNR_{line}$ at the limits for carriers with and without distillation are ~ 24 dB and ~ 15 dB, respectively, both varying by ~ +/- 1 dB. This implies a reduction of the required power per line by 9 dB, suggesting that one could either introduce more comb lines to support a larger overall optical bandwidth of the superchannel transmission, or lower the power of the seeding laser for the EO comb generation to reduce power consumption.

Using the average required $OSNR_{line}$ of 15 dB for the distilled carriers as a benchmark, we fit the curve and interpolated GMI values at this $OSNR_{line}$ for the two cases, to estimate the increase in achievable bit rate achievable by comb distillation. Fig. 7 shows that the values for the narrowband filtered lines lie around the chosen 5% GMI reduction limit of 11.4 b/symbol with a small fluctuation of ≤ 0.1 b/symbol, while for the un-distilled lines GMI values lie around GMI = 9.5 b/symbol with a larger variation of ~ 0.5 b/symbol. This indicates that distillation can improve GMI by about 2 bits/symbol for the 64QAM signals we investigated, showing that comb distillation can provide a real increase in achievable information rates for combs amplified from a low power seed.

In order to qualitatively visualize the impact of comb distillation, Fig. 8 shows the signal constellations for both distilled and un-distilled carriers, at an $OSNR_{line}$ of 15 dB. While it is clear from the constellations that at low $OSNR_{line}$, the signal without distillation through the MRR is noisier (as expected), the magnitude of this noise is higher for outer constellation points, than for the inner ones. This is due to the fact that a noisy carrier was modulated, where noise was added before modulation, and so what we should expect is that the signal amplitude to noise ratio should be static. That means that we should expect a larger noise variance on the higher amplitude (outer) constellation points, as shown.

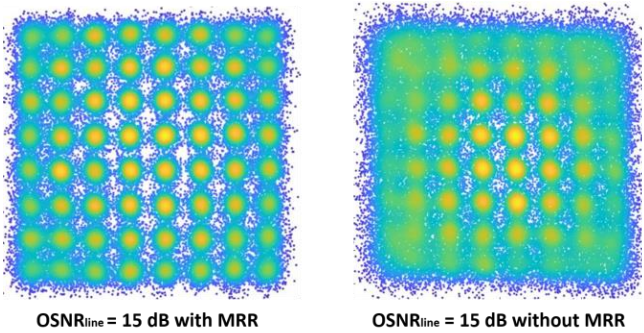

Fig. 8 Signal constellations as a results of distilled and bypassed carrier lines at $OSNR_{line}$ = 15 dB.

## V. DEMONSTRATION OF COMB DISTILLATION FOR LOCAL OSCILLATORS

In this section, we study the impact of comb distillation at the receiver side when applying the same approach as the previous section, i.e. when performing comb filtering for LO lines. In the following experiments, the optical frequency comb was generated at the receiver side, with $OSNR_{line}$ degraded in the same way as the transmitter side experiment, to show the effect of using comb lines that have been amplified from low power as LOs in coherent reception. Similar to the transmitter side experiment, we plotted performance metrics ($Q^2$, BER & GMI) against the $OSNR_{line}$ measured before distillation.

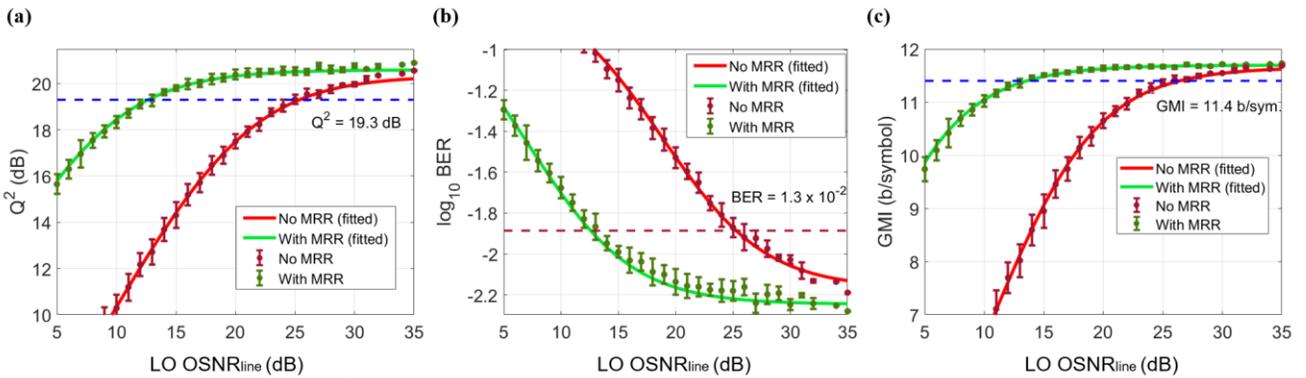

Fig. 9 Performance when noisy and distilled comb lines were employed as LOs in terms of (a) Signal quality ($Q^2$), (b) BER and (c) GMI with dashed lines indicating limits of the 5% reduction of information rates in terms of the particular metrics. Here, the red and green marks indicate undistilled and distilled performance, respectively. The points and error bars show the average and standard deviation of different tested comb lines for specific LO $OSNR_{line}$ values. The red and green lines are guides for the eye for the undistilled and distilled lines, respectively.

*A. Performance in terms of $Q^2$, BER, and GMI*

Fig. 9 (a)-(c) shows performance metrics against comb line OSNR, for the comb lines with and without distillation. The performance traces hit transceiver noise limits at the upper limits of $Q^2 \sim 20$ dB, with a BER $< 10^{-2}$, and a GMI of $\sim 11.6$ b/symbol when the $OSNR_{line}$ reaches $\sim 30$ dB for the un-distilled comb lines (red fitted curves). Again, this changes when filtering with the MRR was applied to the comb lines used as LOs. By inspecting the green fitted curves, $Q^2$ is limited to $\sim 21$ dB for $OSNR_{line}$ of $> 20$ dB, and BER is similarly limited to $\sim 6 \times 10^{-3}$ in the same $OSNR_{line}$ range. GMI reaches a maximum of 11.6 b/symbol at a slightly lower $OSNR_{line}$, not increasing significantly for $OSNR_{line} > 17$ dB. Moreover, after comparing with the un-distilled LO lines, we notice that the LO distillation can contribute to improvements in the $Q^2$ and GMI by up to $\sim 8$ dB and $\sim 3.5$ b/symbol, respectively. This is especially substantial when the optical noise is dominant ($OSNR_{line} \sim 10$ dB).

When considering the 5% reduction in information rate limits, ~22-25 dB $OSNR_{line}$ is required for the bypassed LO lines. These values of $OSNR_{line}$ coincide with the results from the bypassed carriers at the same benchmarks, shown in Fig. 5. This suggests that the required $OSNR_{line}$ is similar at the transmit and receive sides, so a similar limit is seen for per-line and total comb power when distillation is not used.

After comb distillation for LOs, however, the system required only 12-14 dB of $OSNR_{line}$ to attain the 5% reduction in information rate limits. This reduces the required $OSNR_{line}$ for the comb by $\sim 12$ dB compared with the bypassed case. Hence, we conclude that receiver-side LO distillation helps the system achieve the transceiver noise limit at significantly lower $OSNR_{line}$, similarly to the transmitter side.

*B. Performance at the indicative threshold of each comb line*

We next plotted the required $OSNR_{line}$ at a $Q^2 = 19.3$ dB for each tested frequency after interpolating each trace as shown in Fig. 10. We see that the bypassed LO lines possess an overall flat profile of required $OSNR_{line}$ with average required $OSNR_{line}$ at $\sim 25$ dB, conforming to that needed in Fig. 9 (a), with a variation of +/- 1 dB while the distilled LO lines required just 13 dB with the same variation. This translates into a reduction of 12 dB in $OSNR_{line}$, suggesting that distillation could save the per LO line power by up to 15 times.

Comparing the required $OSNR_{line}$ for both carrier and LO distillation cases in Fig. 6 and 10, the narrowband filtering by the MRR seems to be more effective for comb lines deployed as LOs, rather than for carriers, since the $OSNR_{line}$ benefits caused by the MRR can be up to 12 dB for the LOs, while they are 9 dB for the carriers.

To compare the information rate of both bypassed and purified LOs at the chosen benchmark $OSNR_{line} = 13$ dB, we plotted the interpolated GMI at this $OSNR_{line}$, for each line in Fig. 11. Again, we find similar patterns to the transmitter-side experiments where the GMI for the distilled case is around 11.4 b/symbol limit with negligible deviation between individual comb lines. At the $OSNR_{line}$ ($OSNR_{line} = 13$ dB), the GMI for the bypassed LOs is located below $\sim 8$ b/symbol, slightly lower than in the transmitter side case. This is because the required $OSNR_{line}$ of the filtered LOs is 2 dB lower than that of the filtered carriers. This corresponds to the worse performance of the un-distilled LO lines, while the distilled LOs exhibit the similar results to those of distilled carriers. Hence, it is suggested that line distillation at the receiver side is more effective. We infer from this that the beating between LOs and noise when the distilled lines were carriers might be stronger than beating between carriers and noise when the distilled lines were LOs.

We look qualitatively at the effects of LO distillation on the signal constellations. In Fig. 12, in the received signal constellation using a comb line local oscillator with $OSNR_{line} = 13$ dB, we see similar changes in noise distributions with and without distillation to what was observed with a noisy carrier in Fig. 8. Similarly to the case where a noisy carrier was modulated at the transmitter side, when a noisy comb line was used at the receiver, the beating between LO and signal produced a beat amplitude dependent noise variance. As higher signal amplitudes produce a high amplitude beat term in coherent reception, we then expect the high amplitude constellation points to show a higher noise variance than the lower amplitude constellation points, as observed in Fig. 12.

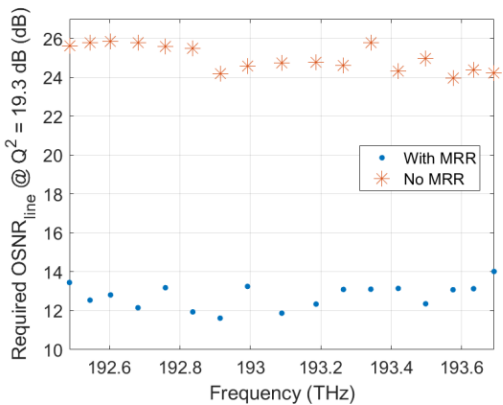

Fig. 10 Required $OSNR_{line}$ at $Q^2 = 19.3$ dB, for both bypassed and distilled LOs against frequency of comb lines.

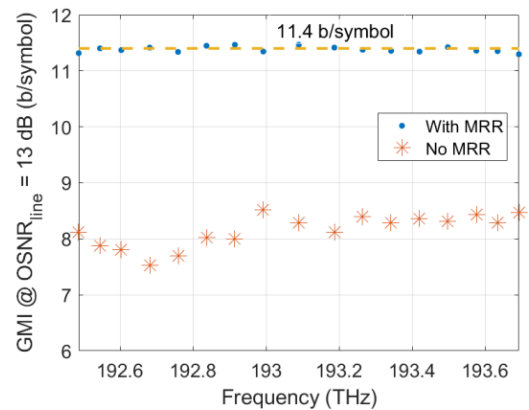

Fig. 11 GMI plot for distilled and bypassed LO lines at $OSNR_{line}$ of 13 dB with the limit of GMI = 11.4 b/symbol as indicated with the dashed line against frequency of comb lines

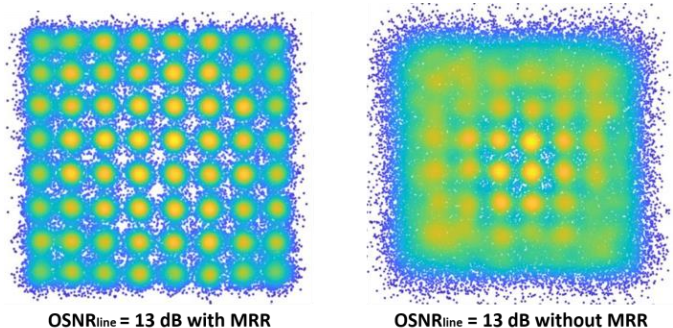

Fig. 12 Signal constellations as a result of distilled and bypassed LO lines at $OSNR_{line}$ = 13 dB.

## VI. Discussion

As we have observed, comb distillation by the MRR offers a reduction in the required $OSNR_{line}$ by ~ 9 and ~ 12 dB for comb lines used as either carriers or LOs, respectively, at a threshold of 5% reduction in information rate, operating at 64QAM. For lower order modulation formats (e.g. QPSK and 16-QAM), we can expect the results to have some difference, due to the lower number of amplitude levels. For QPSK, for example, there is a single amplitude level, and so we expect to see the constellation points all broaden with the same variance. A systematic study on modulation format dependence would highlight if any difference is to be expected. In addition, lower modulation formats have a higher tolerance to noise, and so the $OSNR_{line}$ thresholds, at which significant performance degradations (e.g. reductions in GMI) are observed, are likely to be significantly lower. Conversely, we expect that for higher modulation formats (e.g. 256 QAM and above), that distillation would become important for higher $OSNR_{line}$ ranges than we have investigated here.

While in these experiments we investigated comb distillation for separate cases where the comb lines were used as carriers and LOs, we could also study the system performance where the comb lines were distilled by the MRRs for both the carriers and LOs. This may consist of two independent comb sources and two MRRs, which would seem to be a realistic scenario for comb-based superchannel transmission. We note that not only do the resonance peaks, but also the FSRs of both MRRs, need to strictly match each other to avoid a substantial frequency offset emerging for a significant number of modes relative to the central frequency [17]. By using matched resonators for comb distillation at transmitter and receiver, ones could achieve low power comb (or microcomb) based comb-based high capacity superchannel transmission where effective $OSNR_{line}$ needs to be enhanced. One key system that should be explored, in addition to conventional superchannel transmission, is space-division-multiplexing (SDM) [48], [49] in which power from each comb line is further split to multiple spatial modes. In addition, the practice could also lower power requirement for transmission in short-haul or metro-area links [50].

To lower the optical noise further, higher Q MRRs (Q ~ $10^8$ – $10^9$), which are typically employed in nonlinear optics or microcomb generation [51]–[53], could be employed for comb filtering. Having resonance bandwidths on the order of 100 kHz or so, their filter profiles are close to the linewidths of standard ECLs (linewidth ~ 100 kHz), which may provide a quasi-matched filtering solution for comb distillation. However, this approach would be more sensitive to any frequency mismatch between the resonance peaks and comb lines as a result of any dispersion in the MRRs, thus shifting the resonances from the equidistant grid, as illustrated in Fig. 13. Here, the comb lines match the equidistant grid shown as dashed lines. It is seen that the number of lines undergoing distillation with narrow resonances is limited, since the comb lines cannot reside in the shifted resonance lobes and are thus attenuated by the stopband. This could be, e.g., due to FSR shifts in the MRR due to chromatic dispersion within the ring. Conversely, with a trade-off in in-line noise, distillation by wider resonances would allow more comb lines to transmit through the drop port since the passband regions are wider, and so any detuning of the resonances due to dispersion would have less of an effect. This may be a reason why we do not observe limitations in our comb bandwidth (~1.3 THz) after distillation, with the MRR we use having resonance withs of < 500 MHz.

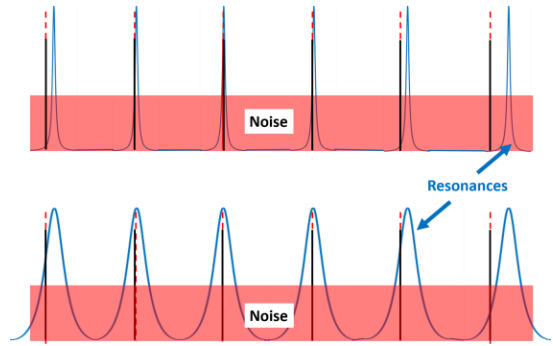

Fig. 13 Comb distillation by the MRR featuring a narrow bandwidth (top) and wide bandwidth (bottom) of resonances, where the comb lines are shown in black with noise floor in red and the equidistant grid is illustrated as the red dashed lines.

Although our approach can lead to substantial noise reduction for comb lines, the comb line spacing must match the fixed FSR of the MRR. This limits the flexibility of the source to be used for transmission at an arbitrary symbol rate. Other potential comb distillation approaches that offer flexibility in line spacing include stimulated Brillouin scattering (SBS), or optical injection locking (OIL).

SBS can be used for comb line purification by using the incoming comb line, and shifting this by the Brillouin frequency shift, then amplifying for use as a pump in a SBS-based amplifier. This then forms a self-tracking narrowband optical amplifier, enabling line filtering with a bandwidth of ~ 30 MHz [32], [33]. When scaling to many parallel lines, the flexibility to variations in line spacing comes at the cost of a high power required for the pump light, and the need for careful consideration of the wavelength dispersion in the SBS shift. Scaling to high comb line counts may require multiple watts of optical power [32], [33], which increases when compensating for dispersion in the SBS shift [54].

For comb purification via optical injection locking (OIL), each comb line is injected into an independent laser, so that the laser output copies the frequency and phase variation of the injected comb line. This approach can also support demultiplexing and amplification simultaneously, by selecting the locking range to be smaller than the line spacing, so that only one line is amplified while the others are discarded [29]. With this principle, it is has been reported [30], [31] that low per-tone power combs can be amplified and used in optical communications systems, achieving better performance than the 'demultiplexing and amplification' method [30], [31]. Similarly, scaling this approach to many parallel lines may have limitations, in that a new laser is required for each line, and each slave laser may also require stabilization to compensate any frequency drifts, e.g., OIL phase lock loop [55] or pilot tone assisted OIL phase locked loop [56].

Reflecting on these alternative techniques, if it is possible to precisely set the comb line spacing in an optical communication system, the high-Q filter distillation approach we investigate here may provide to be desirable in terms of energy efficiency and complexity. If only a small number of lines are required to be distilled, so that the energy consumption and relative complexity of SBS or OIL based approaches is minimal, then the comb spacing need not be precisely set and those approaches may be a preferred alternative.

## VII. Conclusion

We demonstrate an approach to reduce broadband noise centered around frequency comb lines, through the use of filtering with a high Q MRR. We test our approach with an EO comb and find that if the comb line spacing matches the FSR of the MRR, the narrowband filtering by the ring resonances can substantially reduce the optical noise. By benchmarking at a 5% reduction in achievable capacity (GMI), we show that we are able to reduce the required $OSNR_{line}$ for comb lines used as carriers at the transmitter side by 9 dB, and by 12 dB when using comb lines as LOs, through filtering the comb with an MRR. Given this ability to significantly reduce the noise and provide a high degree of parallelism, this suggests that a high Q MRR could be included with low power optical frequency combs (e.g. low-power, chip-scale microcombs) in order to filter in-line noise of a large number amplified of comb lines, to support high bandwidth, high capacity optical communications systems.

## Reference

[1] N. Picqué and T. W. Hänsch, "Frequency comb spectroscopy," *Nat. Photonics*, vol. 13, no. 3, pp. 146–157, 2019, doi: 10.1038/s41566-018-0347-5.

[2] G. Millot *et al.*, "Frequency-agile dual-comb spectroscopy," *Nat. Photonics*, vol. 10, no. 1, pp. 27–30, 2016, doi: 10.1038/nphoton.2015.250.

[3] M. Yu, Y. Okawachi, A. G. Griffith, N. Picqué, M. Lipson, and A. L. Gaeta, "Silicon-chip-based mid-infrared dual-comb spectroscopy," *Nat. Commun.*, vol. 9, no. 1, p. 1869, 2018, doi: 10.1038/s41467-018-04350-1.

[4] J. Feldmann *et al.*, "Parallel convolutional processing using an integrated photonic tensor core," *Nature*, vol. 589, no. 7840, pp. 52–58, 2021, doi: 10.1038/s41586-020-03070-1.

[5] X. Xu *et al.*, "11 TOPS photonic convolutional accelerator for optical neural networks," *Nature*, vol. 589, no. 7840, pp. 44–51, 2021, doi: 10.1038/s41586-020-03063-0.

[6] M. Mazur, A. Lorences-Riesgo, J. Schröder, P. A. Andrekson, and M. Karlsson, "High Spectral Efficiency PM-128QAM Comb-Based Superchannel Transmission Enabled by a Single Shared Optical Pilot Tone," *J. Light. Technol.*, vol. 36, no. 6, pp. 1318–1325, Mar. 2018, doi: 10.1109/JLT.2017.2786750.

[7] A. Pasquazi *et al.*, "Micro-combs: A novel generation of optical sources," *Phys. Rep.*, vol. 729, pp. 1–81, 2018.

[8] A. L. Gaeta, M. Lipson, and T. J. Kippenberg, "Photonic-chip-based frequency combs," *Nat. Photonics*, vol. 13, no. 3, pp. 158–169, 2019, doi: 10.1038/s41566-019-0358-x.

[9] T. J. Kippenberg, A. L. Gaeta, M. Lipson, and M. L. Gorodetsky, "Dissipative Kerr solitons in optical microresonators," *Science (80-. ).*, vol. 361, no. 6402, 2018, doi: 10.1126/science.aan8083.

[10] X. Xu, M. Tan, J. Wu, R. Morandotti, A. Mitchell, and D. J. Moss, "Microcomb-Based Photonic RF Signal Processing," *IEEE Photonics Technol. Lett.*, vol. 31, no. 23, pp. 1854–1857, Dec. 2019, doi: 10.1109/LPT.2019.2940497.

[11] J. Wu *et al.*, "RF Photonics: An Optical Microcombs' Perspective," *IEEE J. Sel. Top. Quantum Electron.*, vol. 24, no. 4, pp. 1–20, Jul. 2018, doi: 10.1109/JSTQE.2018.2805814.

[12] P. Marin-Palomo *et al.*, "Microresonator-based solitons for massively parallel coherent optical communications," *Nature*, vol. 546, no. 7657, pp. 274–279, 2017, doi: 10.1038/nature22387.

[13] J. Pfeifle *et al.*, "Optimally Coherent Kerr Combs Generated with Crystalline Whispering Gallery Mode Resonators for Ultrahigh Capacity Fiber Communications," *Phys. Rev. Lett.*, vol. 114, no. 9, p. 93902, Mar. 2015, doi: 10.1103/PhysRevLett.114.093902.

[14] B. Corcoran *et al.*, "Ultra-dense optical data transmission over standard fibre with a single chip source," *Nat. Commun.*, vol. 11, no. 1, p. 2568, 2020, doi: 10.1038/s41467-020-16265-x.

[15] S. Chandrasekhar, Xiang Liu, B. Zhu, and D. W. Peckham, "Transmission of a 1.2-Tb/s 24-carrier no-guard-interval coherent OFDM superchannel over 7200-km of ultra-large-area fiber," in *2009 35th European Conference on Optical Communication*, 2009, vol. 2009-Suppl, pp. 1–2.

[16] J. Renaudier, R. R. Müller, L. Schmalen, P. Tran, P. Brindel, and G. Charlet, "1-Tb/s PDM-32QAM superchannel transmission at 6.7-b/s/Hz over SSMF and 150-GHz-grid ROADMs," in *2014 The European Conference on Optical Communication (ECOC)*, 2014, pp. 1–3.

[17] V. Torres-Company *et al.*, "Laser Frequency Combs for Coherent Optical Communications," *J. Light. Technol.*, vol. 37, no. 7, pp. 1663–1670, Apr. 2019, [Online]. Available: http://jlt.osa.org/abstract.cfm?URI=jlt-37-7-1663.

[18] M. Mazur, A. Lorences-Riesgo, J. Schröder, P. A. Andrekson, and M. Karlsson, "10 Tb/s PM-64QAM self-homodyne comb-based superchannel transmission with 4% shared pilot tone overhead," *J. Light. Technol.*, vol. 36, no. 16, pp. 3176–3184, 2018.

[19] L. Lundberg *et al.*, "Phase-coherent lightwave communications with frequency combs," *Nat. Commun.*, vol. 11, no. 1, p. 201, 2020, doi: 10.1038/s41467-019-14010-7.

[20] T. Herr *et al.*, "Universal formation dynamics and noise of Kerr-frequency combs in microresonators," *Nat. Photonics*, vol. 6, no. 7, pp. 480–487, 2012, doi: 10.1038/nphoton.2012.127.

[21] S. J. Herr *et al.*, "Frequency comb up- and down-conversion in synchronously driven $\chi^2$ optical microresonators," *Opt. Lett.*, vol. 43, no. 23, pp. 5745–5748, Dec. 2018, doi: 10.1364/OL.43.005745.

[22] C. R. Phillips *et al.*, "Supercontinuum generation in quasi-phase-matched LiNbO3 waveguide pumped by a Tm-doped fiber laser system," *Opt. Lett.*, vol. 36, no. 19, pp. 3912–3914, 2011, doi: 10.1364/OL.36.003912.

[23] Y. Huang *et al.*, "Temporal soliton and optical frequency comb generation in a Brillouin laser cavity," *Optica*, vol. 6, no. 12, pp. 1491–1497, 2019, doi: 10.1364/OPTICA.6.001491.

[24] M. Zhang *et al.*, "Broadband electro-optic frequency comb generation in a lithium niobate microring resonator," *Nature*, vol. 568, no. 7752, pp. 373–377, 2019, doi: 10.1038/s41586-019-1008-7.

[25] A. J. Metcalf, V. Torres-Company, D. E. Leaird, and A. M. Weiner,

"High-Power Broadly Tunable Electrooptic Frequency Comb Generator," *IEEE J. Sel. Top. Quantum Electron.*, vol. 19, no. 6, pp. 231–236, Nov. 2013, doi: 10.1109/JSTQE.2013.2268384.

[26] M. Kourogi, K. Nakagawa, and M. Ohtsu, "Wide-span optical frequency comb generator for accurate optical frequency difference measurement," *IEEE J. Quantum Electron.*, vol. 29, no. 10, pp. 2693–2701, Oct. 1993, doi: 10.1109/3.250392.

[27] A. Pasquazi, M. Peccianti, B. E. Little, S. T. Chu, D. J. Moss, and R. Morandotti, "Stable, dual mode, high repetition rate mode-locked laser based on a microring resonator," *Opt. Express*, vol. 20, no. 24, pp. 27355–27363, 2012, doi: 10.1364/OE.20.027355.

[28] Z. Y. Zhang et al., "1.55 μm InAs/GaAs Quantum Dots and High Repetition Rate Quantum Dot SESAM Mode-locked Laser," *Sci. Rep.*, vol. 2, no. 1, p. 477, 2012, doi: 10.1038/srep00477.

[29] Z. Liu and R. Slavík, "Optical Injection Locking: From Principle to Applications," *J. Light. Technol.*, vol. 38, no. 1, pp. 43–59, Jan. 2020, doi: 10.1109/JLT.2019.2945718.

[30] A. Albores-Mejia, T. Kaneko, E. Banno, K. Uesaka, H. Shoji, and H. Kuwatsuka, "Optical-Comb-Line Selection from a Low-Power/Low-OSNR Comb using a Low-Coherence Semiconductor Laser for Flexible Ultra-Dense Short Range Transceivers," in *Optical Fiber Communication Conference*, 2015, p. W2A.23, doi: 10.1364/OFC.2015.W2A.23.

[31] R. Zhou, T. Shao, M. D. G. Pascual, F. Smyth, and L. P. Barry, "Injection Locked Wavelength De-Multiplexer for Optical Comb-Based Nyquist WDM System," *IEEE Photonics Technol. Lett.*, vol. 27, no. 24, pp. 2595–2598, 2015, doi: 10.1109/LPT.2015.2478791.

[32] A. Choudhary et al., "On-chip Brillouin purification for frequency comb-based coherent optical communications," *Opt. Lett.*, vol. 42, no. 24, pp. 5074–5077, Dec. 2017, doi: 10.1364/OL.42.005074.

[33] M. Pelusi et al., "Low noise frequency comb carriers for 64-QAM via a Brillouin comb amplifier," *Opt. Express*, vol. 25, no. 15, pp. 17847–17863, Jul. 2017, doi: 10.1364/OE.25.017847.

[34] B. Corcoran et al., "Overcoming low-power limitations on optical frequency combs using a micro-ring resonator," in *Optical Fiber Communication Conference (OFC) 2020*, 2020, p. T4G.5, doi: 10.1364/OFC.2020.T4G.5.

[35] K. Beha, D. C. Cole, P. Del'Haye, A. Coillet, S. A. Diddams, and S. B. Papp, "Electronic synthesis of light," *Optica*, vol. 4, no. 4, pp. 406–411, Apr. 2017, doi: 10.1364/OPTICA.4.000406.

[36] W. Bogaerts et al., "Silicon microring resonators," *Laser Photon. Rev.*, vol. 6, no. 1, pp. 47–73, Jan. 2012, doi: https://doi.org/10.1002/lpor.201100017.

[37] D. J. Moss, R. Morandotti, A. L. Gaeta, and M. Lipson, "New CMOS-compatible platforms based on silicon nitride and Hydex for nonlinear optics," *Nat. Photonics*, vol. 7, no. 8, pp. 597–607, 2013, doi: 10.1038/nphoton.2013.183.

[38] M. Ferrera et al., "Low-power continuous-wave nonlinear optics in doped silica glass integrated waveguide structures," *Nat. Photonics*, vol. 2, no. 12, pp. 737–740, 2008, doi: 10.1038/nphoton.2008.228.

[39] B. Stern, X. Ji, Y. Okawachi, A. L. Gaeta, and M. Lipson, "Battery-operated integrated frequency comb generator," *Nature*, vol. 562, no. 7727, pp. 401–405, 2018, doi: 10.1038/s41586-018-0598-9.

[40] M. Mazur et al., "High Spectral Efficiency Coherent Superchannel Transmission With Soliton Microcombs," *J. Light. Technol.*, vol. 39, no. 13, pp. 4367–4373, 2021, doi: 10.1109/JLT.2021.3073567.

[41] K. Kikushima and H. Yoshinaga, "Distortion due to gain tilt of erbium-doped fiber amplifiers," *IEEE Photonics Technol. Lett.*, vol. 3, no. 10, pp. 945–947, Oct. 1991, doi: 10.1109/68.93272.

[42] M. S. Faruk and S. J. Savory, "Digital Signal Processing for Coherent Transceivers Employing Multilevel Formats," *J. Light. Technol.*, vol. 35, no. 5, pp. 1125–1141, Mar. 2017, [Online]. Available: http://jlt.osa.org/abstract.cfm?URI=jlt-35-5-1125.

[43] K. Kikuchi, "Fundamentals of Coherent Optical Fiber Communications," *J. Light. Technol.*, vol. 34, no. 1, pp. 157–179, Jan. 2016, doi: 10.1109/JLT.2015.2463719.

[44] Y. Mori, C. Zhang, and K. Kikuchi, "Novel configuration of finite-impulse-response filters tolerant to carrier-phase fluctuations in digital coherent optical receivers for higher-order quadrature amplitude modulation signals," *Opt. Express*, vol. 20, no. 24, pp. 26236–26251, 2012.

[45] Y. Gao, A. P. T. Lau, S. Yan, and C. Lu, "Low-complexity and phase noise tolerant carrier phase estimation for dual-polarization 16-QAM systems," *Opt. Express*, vol. 19, no. 22, pp. 21717–21729, 2011, doi: 10.1364/OE.19.021717.

[46] R. A. Shafik, M. S. Rahman, and A. R. Islam, "On the Extended Relationships Among EVM, BER and SNR as Performance Metrics," in *2006 International Conference on Electrical and Computer Engineering*, Dec. 2006, pp. 408–411, doi: 10.1109/ICECE.2006.355657.

[47] T. Fehenberger, "calcGMIscript.m.", Apr. 2015, Accessed on: Mar. 06, 2021, [Online]. Available: https://www.fehenberger.de/#sourcecode.

[48] G. Rademacher et al., "10.66 Peta-Bit/s Transmission over a 38-Core-Three-Mode Fiber," in *Optical Fiber Communication Conference (OFC) 2020*, 2020, p. Th3H.1, doi: 10.1364/OFC.2020.Th3H.1.

[49] B. J. Puttnam et al., "2.15 Pb/s transmission using a 22 core homogeneous single-mode multi-core fiber and wideband optical comb," in *2015 European Conference on Optical Communication (ECOC)*, 2015, pp. 1–3, doi: 10.1109/ECOC.2015.7341685.

[50] P. Marin-Palomo, J. N. Kemal, T. J. Kippenberg, W. Freude, S. Randel, and C. Koos, "Performance of chip-scale optical frequency comb generators in coherent WDM communications," *Opt. Express*, vol. 28, no. 9, pp. 12897–12910, 2020, doi: 10.1364/OE.380413.

[51] T. Herr et al., "Temporal solitons in optical microresonators," *Nat. Photonics*, vol. 8, no. 2, pp. 145–152, 2014, doi: 10.1038/nphoton.2013.343.

[52] P. Del'Haye, S. A. Diddams, and S. B. Papp, "Laser-machined ultra-high-Q microrod resonators for nonlinear optics," *Appl. Phys. Lett.*, vol. 102, no. 22, p. 221119, 2013, doi: 10.1063/1.4809781.

[53] T. J. Kippenberg, S. M. Spillane, and K. J. Vahala, "Kerr-Nonlinearity Optical Parametric Oscillation in an Ultrahigh-$Q$ Toroid Microcavity," *Phys. Rev. Lett.*, vol. 93, no. 8, p. 83904, Aug. 2004, doi: 10.1103/PhysRevLett.93.083904.

[54] A. Zarifi, M. Merklein, Y. Liu, A. Choudhary, B. J. Eggleton, and B. Corcoran, "EDFA-band Coverage Broadband SBS Filter for Optical Carrier Recovery," in *2020 Conference on Lasers and Electro-Optics Pacific Rim (CLEO-PR)*, Aug. 2020, pp. 1–2, doi: 10.1364/CLEOPR.2020.C9G_3.

[55] A. C. Bordonalli, C. Walton, and A. J. Seeds, "High-Performance Phase Locking of Wide Linewidth Semiconductor Lasers by Combined Use of Optical Injection Locking and Optical Phase-Lock Loop," *J. Light. Technol.*, vol. 17, no. 2, p. 328, Feb. 1999, [Online]. Available: http://jlt.osa.org/abstract.cfm?URI=jlt-17-2-328.

[56] Z. Liu, J. Kim, D. S. Wu, D. J. Richardson, and R. Slavík, "Homodyne OFDM with Optical Injection Locking for Carrier Recovery," *J. Light. Technol.*, vol. 33, no. 1, pp. 34–41, Jan. 2015, doi: 10.1109/JLT.2014.2369994.